Costantino Sigismondi
*Università di Roma*
*"La Sapienza"*
sigismondi@icra.it

# Misura del diametro solare ad almucantarat zero

**Abstract:** Among several methods for accurate measurements of solar diameter, there are transits at fixed almucantarat used in solar astrolabes. The almucantarat is a circle of given height. The horizon circle is the zero almucantarat, and data from four sunsets on the Tyrrenian sea are discussed. The angular diameter of the Sun is recovered with a few arcseconds accuracy using 60 fps video of sunsets.

### Introduzione: diametro ed almucantarat, gli astrolabi solari

Si vuol mettere in evidenza che i principi su cui si basano gli astrolabi solari sono applicabili anche ai tramonti sull'orizzonte marino (intento didattico) e possono fornire risultati che-opportunamente trattati diventano interessanti di per sé e per il metodo di analisi dati per essi sviluppato (come nel trattamento di un effetto di tipo "black drop" riutilizzabile per i futuri transiti planetari sul Sole). L'idea di misurare il diametro del Sole cronometrando il transito attraverso un cerchio di uguale altezza [1] è stata messa in opera realizzando i cosiddetti astrolabi solari. Questi astrolabi, dal punto di vista etimologico, sono strumenti atti a "prendere" le dimensioni del Sole, nella fattispecie. L'inventore di questi strumenti, usati per l'astonomia di posizione, è stato André Danjon [2], [3].

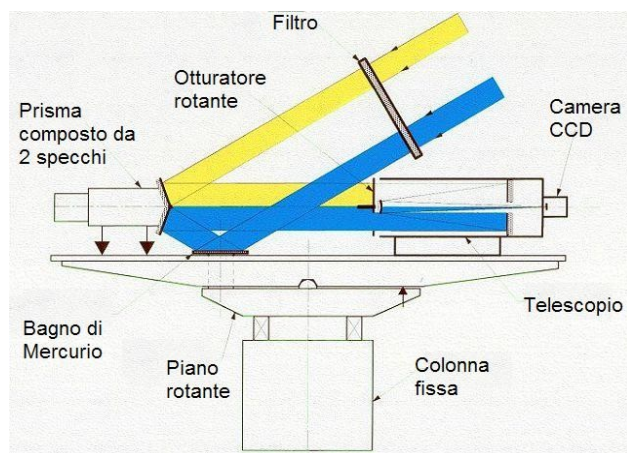

Figura 1. Schema ottico dell'astrolabio DORAYSOL, Définition et Observation du RAYon SOLaire (esperimento francese operativo dal 1990 al 2008) [4].

DORAYSOL è l'ultima generazione degli astrolabi solari, altrimenti muniti di una serie di prismi intercambiabili al posto dei due specchi. L'inclinazione degli specchi è cambiata con continuità da un motore apposito per servire per tutte le distanze zenitali z entro cui lavora lo strumento (30°<z<60°). Il bagno di mercurio costituisce lo specchio di garantita orizzontalità. Al mattino l'immagine diretta (gialla) sale, mentre la riflessa (blu) scende.
Un otturatore rotante consente di formare alternativamente l'immagine diretta e quella riflessa sulla camera.

Il loro incrociarsi determina gli istanti di prima t1 e seconda t2 tangenza con l'almucantarat (linea orizzontale in figura 2) di data altezza.

**Almucantarat zero con immagini diretta e riflessa**
L'orizzonte del mare è sicuramente orizzontale. Con una ripresa video a 60 fps si possono riprendere con 1/60s di precisione t1 e t2. Data la velocità di rotazione della sfera celeste di circa 15"/s, 1/60 s corrisponde a 0.25" di accuratezza angolare nominale.

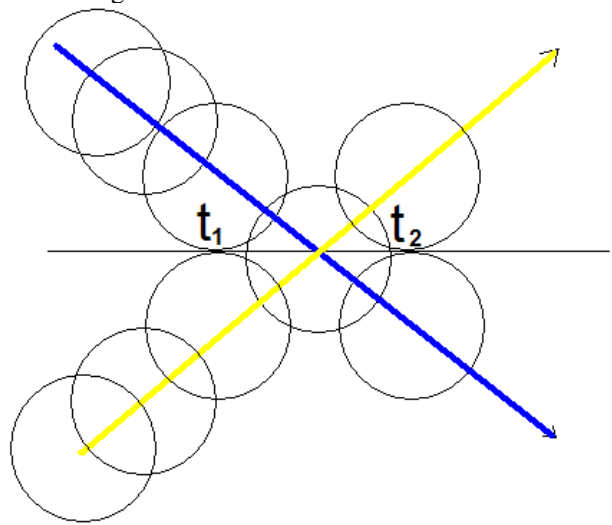

Figura 2. Immagini diretta (gialla) e riflessa (blu) al mattino, riprese da DORAYSOL. L'intervallo di tempo t2-t1 è proporzionale al diametro verticale del Sole.

Al tramonto, analogamente con DORAYSOL, si può osservare anche l'immagine solare riflessa ed il contatto tra le due immagini dà t1. Il tempo t2 è l'ultimo bagliore di Sole, essendo un segnale di tipo SI/NO, questo è quello determinato in modo più preciso ± 1 frame cioè ±1/60s.

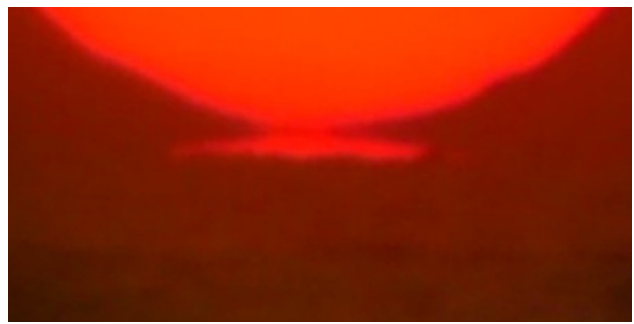

Figura 3. Primo contatto con l'orizzonte al tramonto del 13 agosto 2009 da Ostia (Idroscalo, foce del Tevere). L'immagine è estratta da un video a 60 fps, preso con una videocamera Sanyo CG9 a 5x posta dietro un monocolo 8x21, la sensibilità è stata impostata a 50 ISO. Si noti, al centro, l'effetto "black drop" di anticipato collegamento luminoso tra le due immagini diretta e riflessa, risultato della combinazione tra ottica strumentale e funzione di

luminosità del Sole, che cambia rapidamente d'intensità nei pressi del bordo [8].

## Trasmittanza e proprietà dell'atmosfera lungo la linea di vista all'orizzonte

L'ultimo contatto t2 è un segnale SI/NO simile alla sparizione di un grano di Baily durante un'eclissi solare [5], e non è influenzato dal seeing atmosferico, ma solo dal valore della rifrazione all'orizzonte, che si assume costante durante tutto il tempo del tramonto $\Delta t = t2-t1$, poiché è una proprietà integrale lungo tutta la linea di vista che all'orizzonte intercetta più di 5 masse d'aria. [6]

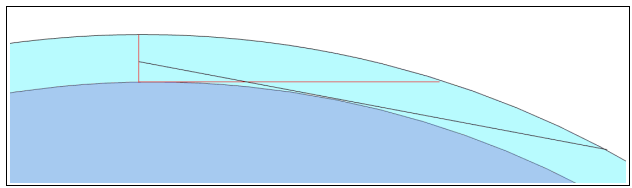

Figura 4. Modello di estinzione atmosferica: lo spessore è tale che all'orizzonte ed al livello del mare la linea di vista (in rosso) attraversa 5 volte la quantità d'aria che sovrasta lo zenith. Se si osserva da una posizione elevata h [m] sul livello del mare, la quantità d'aria intercettata aumenta di $\Delta m \approx 0.36\sqrt{h}$. Quindi, come esempio, osservando da una quota di h=200 m abbiamo un'aggiunta di $\Delta m \approx 5$ masse d'aria guardando verso l'orizzonte visibile (linea nera).

Tuttavia la foschia è la principale responsabile del calo di luminosità del Sole presso l'orizzonte, e questa corrisponde all'effetto di uno strato (cfr. fig. 5) di circa 10-30 m: questo strato influisce poco sulla trasmittanza atmosferica a piccole distanze zenitali, ma diventa importante presso l'orizzonte, determinando facilmente 10 magnitudini di caduta dell'intensità; il suo effetto dipende dalla trasparenza del cielo di ogni sera, che è estremamente variabile, e poiché [7] $I/I_0 \approx 0.855^{(m)}$, con m il numero di masse d'aria, la foschia nella direzione dell'orizzonte corrisponde all'effetto di 60 masse d'aria!

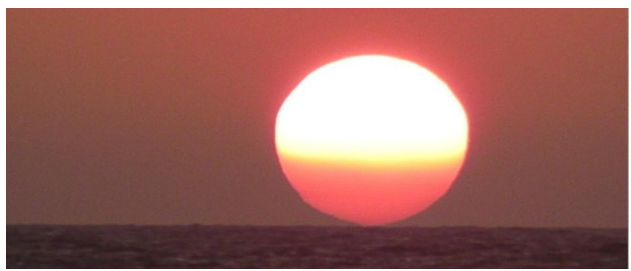

Figura 5. La parte inferiore del Sole (circa 15 arcominuti) è molto meno intensa della parte superiore, che satura il rivelatore. 15' alla distanza dell'orizzonte visto in questa immagine da 4 m di quota, sono 30 m. Lo strato di foschia in quel giorno era attorno ai 10 metri di spessore. Ostia 11.8.2009, immagine estratta da video a 60 fps preso con videocamera Sanyo HD10 impostata a 50 ISO e zoom 50 x.

## Definizione del tempo di primo contatto con l'orizzonte

In definitiva tutto il problema della misura del diametro solare all'orizzonte sta nella determinazione dell'istante di primo contatto del Sole con l'orizzonte.
Nel caso del tramonto, la funzione $f(t)$ è l'intersezione tra il disco solare che scende e la linea dell'orizzonte.

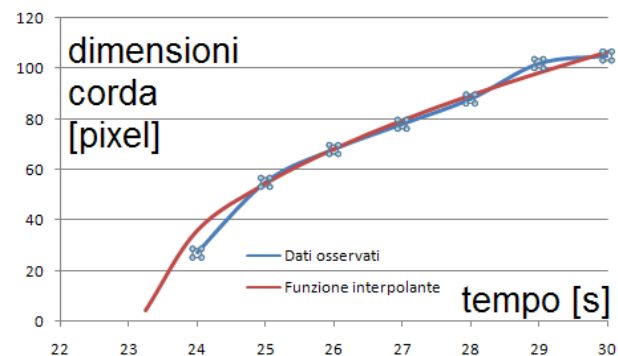

Figura 6. La funzione interpolante $f(t)$ calcola lunghezza della corda tagliata sul disco solare dall'orizzonte: è una radice quadrata, che minimizza il $\chi^2$ dei dati osservati.

Al valore corrispondente a dimensione nulla della corda si associa un'incertezza nominale data da $\Delta\chi^2=1$. Nei dati del tramonto dell'11 agosto 2009 [in fig. 5] abbiamo t1=23.23±0.04 s, con un $\chi^2/\nu \approx 2.9$ ($\nu$=7).

## Analisi dati

L'osservazione il 15.7.09 è fatta ad occhio nudo, il 9.8.09 col binocolo 8x21ed il tramonto era con riflesso, l'11.8 video del tramonto senza riflesso, mentre il 13.8 con riflesso. Nella tabella seguente si riassumono i risultati; solo i due video sono di fatto *diffraction limited*; la diffrazione qui vale 5" ed allarga in modo sistematico il diametro.

| Data | $\Delta t$ calc [s] | $\Delta t$ oss [s] | $\Delta\emptyset$ ["] |
|---|---|---|---|
| 15.7.09 occhio | 194.19 | 204±0.4 | +137±6 |
| 9.8.09 binocolo | 182.0 | 182±0.4 | +0±6 |
| 11.8.09 video | 180.64 | 180.95±0.05 | +4.5±0.8 |
| 13.8.09 video | 180.02 | 180.10±0.05 | +1.1±0.8 |

Tabella 1. La tabella indica le correzioni $\Delta\emptyset$ ["]-delta diametro- riportate rispetto al diametro solare standard 1919.26". Così, ad esempio, il 15 luglio 2009 il diametro del Sole valeva 1888.26" ed avrebbe comportato un tramonto di durata 194.19 s; l'osservazione di un tramonto di 204±0.4 s implica un rapporto tra diametro teorico e diametro osservato pari a quello tra le durate calcolata ed osservata. Ovvero un incremento percentuale di (204-194.19)/194.19, che moltiplicato per il diametro standard 1919.26" ridà $\Delta\emptyset$=137". Analogamente l'incertezza Lo stesso procedimento è stato fatto per gli altri 3 tramonti. L'uso di riferirsi al diametro angolare standard del Sole mostra l'interesse verso le variazioni angolari intrinseche al Sole e non a quelle determinate dalla normale variazione stagionale della distanza Terra-Sole.

Si vede che pure con questi strumenti relativamente modesti (con obbiettivi di circa 2cm di diametro) quanto a risoluzione angolare si ottengono dei valori estremamente interessanti, che convergono a zero all'aumentare della precisione del metodo. La durata del tramonto è calcolata col programma Ephemvga [9] inserendo le coordinate del luogo di osservazione, in assenza di atmosfera e sull'almucantarat corrispondente all'istante di sparizione del Sole. La rifrazione infatti modifica l'altezza apparente dell'astro [10], ma non il tempo di transito su un dato almucantarat.

**Discussione**

La videocamera deve restare fissa durante tutto il tramonto altrimenti uno spostamento di 3 cm in verticale corrisponde alla distanza dell'orizzonte di 7 Km a circa 0.9".

La luminosità del Sole, fino alla tangenza, limita la precisione della misura di t1. Se non c'è abbastanza foschia è utile filtrare il Sole con 2 polarizzatori rotanti che riducono l'intensità luminosa con continuità secondo la legge di Malus. Le videocamere commerciali hanno degli algoritmi che tendono a simulare la risposta dell'occhio umano, nel caso di forte contrasto luce-buio come per il lembo solare, questo è sensibilmente spostato verso il buio. Insieme alla diffrazione questo effetto tende ad anticipare la misura del tempo t1. L'interpolazione con la funzione *f(t)* della corda riduce questo effetto.

La corda comunque è definita tra i primi pixel non illuminati, laddove la definizione classica di lembo solare [11] è il massimo della derivata della curva di luce.

Il valore migliore del diametro è stato quello ottenuto con il tramonto con riflesso ed il Sole con diametro di 300-400 pixel. Questo risultato con obbiettivi di 2 cm di diametro incoraggia a proseguire con telescopi maggiori.

Le misure di diametro solare del satellite SOHO, calibrato per questo scopo con il transito di Mercurio del 7.5.2003[12] sono ottenute a partire dai pixel (dimensione dell'immagine); mentre nei metodi con transiti (su cerchi orari o di altezza) è il tempo che fornisce la misura del diametro.

Nel caso qui esaminato il metodo è certamente accessibile a tutti, e se fatto con telescopi di 10-20 cm promette ottimi risultati.